\documentclass[numbers,sort&compress]{IntechOpen-Book}

\usepackage{wrapfig}

\usepackage{ragged2e}

\usepackage{tikz}
\usepackage{xcolor}
\newcommand*\circled[1]{\tikz[baseline=(char.base)]{
            \node[shape=circle,fill,inner sep=2pt] (char) {\textcolor{white}{#1}};}}
            
\usepackage{multirow}
\newcommand{\pfault}{\textsc{PFault}}

\graphicspath{{Artworks/}}

\begin{document}
\justifying

\Mainmatter

\begin{frontmatter}

\chapter{On Fault Tolerance of Data Storage Systems: A Holistic Perspective}

 \author{Mai Zheng}
\author{Duo Zhang}
 \author{Ahmed Dajani}

\makechaptertitle

\chaptermark{On Fault Tolerance of Data Storage Systems: A Holistic Perspective}

\begin{abstract}
Data storage systems serve as the foundation of  digital society. The enormous  data generated by people on a daily basis make the fault tolerance of data storage systems increasingly important. Unfortunately, modern storage systems consist of complicated hardware and software layers interacting with each other, which may contain latent bugs that elude extensive testing and lead to data corruption, system downtime, or even unrecoverable data loss in practice.
 In this chapter, we take a holistic view to introduce the typical architecture and major components of modern data storage systems (e.g., solid state drives, persistent memories,  local file systems, and distributed storage management  at scale). Next, we discuss a few representative bug detection and fault tolerance techniques across layers with a focus on issues that affect system recovery and data integrity. Finally, we conclude with open challenges and future work.

\end{abstract}

\begin{keywords} 
Data Storage, File System, System Failure, Data Loss, Metadata Corruption, Crash Consistency, Fault Injection, Fault Tolerance, Data Integrity, Reliability, Resilience, Security 
\end{keywords}

\end{frontmatter}

\section{Introduction}
\label{sec:intro}


Data storage systems play an essential role in modern digital society. The enormous data generated by various use cases   (e.g.,  financial transactions, medical records, scientific datasets) continuously make the fault tolerance of data storage systems increasingly important. 
Unfortunately, building a fault-tolerant storage system is 
challenging  due to the ever-growing complexity. Modern storage systems consist of complicated hardware and software layers interacting with each other, which may contain latent bugs that elude extensive testing and hurt the data integrity once triggered.
In particular, latent defects in fault-tolerance mechanisms can lead to severe consequences, including server downtime, data corruption, and  financial losses. 
As data storage systems continue to evolve and grow in terms of scale and complexity, the risk of failures becomes increasingly prevalent~\cite{hothardware20201220,algolia_blog,algolia-samsung-patch}. Therefore, understanding the system architecture and analyzing the fault tolerance thoroughly is imperative. 

\begin{figure*}[htb] 
    \begin{center} 
    \includegraphics[width=4.8in]{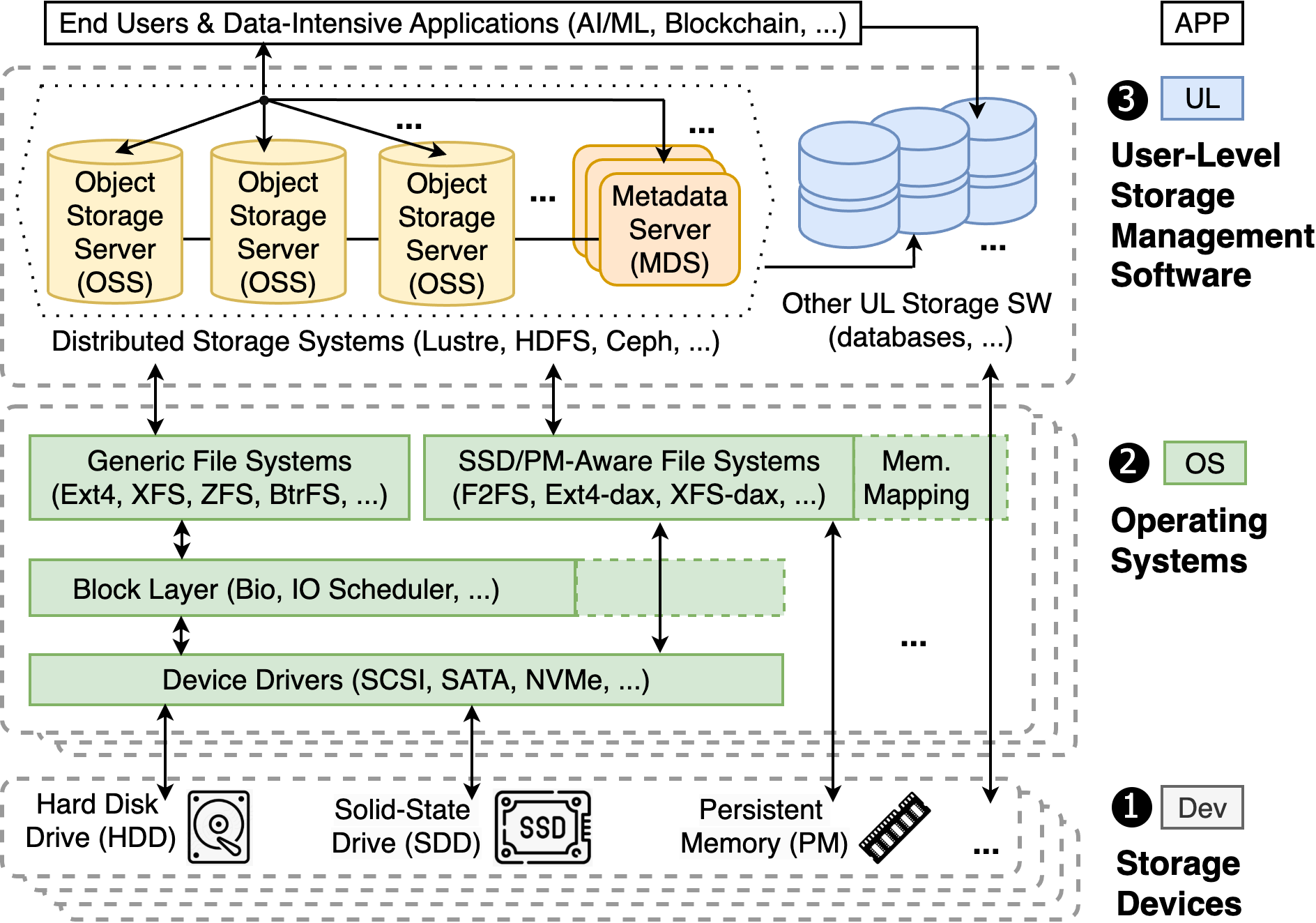} 
    \end{center} 
    \vspace{-0.2in}
    \caption{\bf A Holistic View of Data Storage Systems.} 
    \label{fig:stack} 
\end{figure*}

In this section, we introduce the typical architecture of   modern data storage systems  to lay the foundation for further discussions on storage fault tolerance (\S\ref{sec:frameworks}). As shown in Figure \ref{fig:stack}, a typical data storage system mainly consists of three logical layers: storage devices (\circled{1} Dev), operating systems (\circled{2} OS), and user-level storage management software (\circled{3} UL). Each of them plays a unique role in managing data in computers, and the combination of them provides  end-to-end  supports from  interacting with hardware to handling  user requests for various data-intensive application scenarios (e.g., blockchain transactions, artificial intelligence (AI) and machine learning (ML) applications). We elaborate on the three logical layers from the bottom up in the following subsections.

\subsection{Storage Devices (Dev): Persistent Home for Data}

Storage devices serve as the foundation of data storage systems to provide the necessary persistency to store data durably. 
There are various types of storage devices with different tradeoffs in terms of performance, cost, etc. today, and we briefly introduce three main ones below:

\smallskip
\noindent
\textbf{Hard Disk Drives (HDD).} 
HDDs have been the dominant storage medium for decades due to their cost-effectiveness and high capacity. HDDs use magnetic platters to store data, with a mechanical arm positioning a read/write head over spinning disks to access information. Despite their affordability and long lifespan, HDDs suffer from relatively slow access speeds due to mechanical latency and seek time. Typical rotational speeds range from 5,400 to 15,000 RPM (revolutions per minute) today, leading to access times in milliseconds. HDDs are particularly well-suited for archival and bulk storage applications where cost per gigabyte is a priority. However, they struggle to meet the performance demands of modern workloads that require fast random access. To mitigate their limitations, techniques like caching and shingled magnetic recording (SMR) have been introduced
to improve HDDs in terms of performance and/or capacity.

\smallskip
\noindent
\textbf{Flash-based Solid-State Drives (SSD).} SSDs have revolutionized storage by eliminating mechanical components, instead relying on NAND flash memory to store data electronically. Compared to HDDs, SSDs offer much lower latencies, with typical access times in the microsecond range. The absence of moving parts enables SSDs to deliver high-speed random reads and writes, making them ideal for performance-sensitive applications. SSDs come in various form factors with NVMe-based drives offering significantly higher bandwidth than other traditional counterparts. However, NAND flash memory has inherent limitations, including limited write endurance and higher cost per byte compared to HDDs. 
To mitigate wear, SSDs employ techniques like wear leveling and garbage collection, which may introduce write amplification overhead. The transition from planar NAND to 3D NAND has improved density and endurance, helping SSDs achieve wider adoption across consumer and enterprise markets.

\smallskip
\noindent
\textbf{Persistent Memory (PM).} PM technologies offer attractive features for developing storage systems and applications. Unlike traditional volatile memory technology (i.e., DRAM), PM provides non-volatility, allowing data to persist across power cycles while maintaining low-latency memory access. For example, Intel\textsuperscript{\textregistered} Optane\textsuperscript{\texttrademark}~\cite{PM} can support byte-granularity accesses with latencies less than $3\times$ of DRAM latencies~\cite{yang2020empirical}, while also providing durability guarantees. These properties enable PM to serve as both a high-speed storage tier and an extension of main memory, bridging the performance gap between DRAM and SSDs. 
However, PM also introduces new challenges, such as wear endurance limitations and the need for efficient software interfaces to leverage its byte-addressable capabilities. 
As persistent memory technologies continue to evolve, they hold promise for reshaping storage hierarchies and enabling new classes of high-performance applications.

Note that it is possible to organize multiple storage devices in an array to improve parallelism and redundancy, so as to achieve higher performance and/or fault tolerance (e.g., RAID~\cite{raid88}).
In general, these storage devices including RAID-like solutions exhibit different characteristics and failure modes, which demands different strategies for ensuring fault tolerance in storage systems.

\subsection{Storage Software Stack in OS: Managing Data on A Single Computer}

The storage software stack in the operating system (OS)   serves as an intermediary between user applications and physical  devices to manage data on a single computer.
We briefly introduce a few key components in the widely used Linux OS below, and refer the interested readers to~\cite{love2010linux,ostep} for more details.

\smallskip
\noindent
\textbf{Device Drivers.}The device drivers directly communicate with storage devices based on standardized interfaces (e.g., SCSI, SATA, NVMe). Each interface defines a set of commands and protocols for communications. 
In particular, modern NVMe drivers leverage multi-queue capabilities to maximize parallelism and reduce latency, significantly improving SSD performance. Persistent memory devices rely on drivers like Linux’s \texttt{libnvdimm} to manage NVDIMM devices and expose them as either block devices or memory-mapped regions. Efficient driver implementation is crucial for ensuring low-latency, high-throughput storage access, particularly for emerging technologies like CXL-attached memory and computational storage.

\smallskip
\noindent
\textbf{The Block I/O Layer.}The block layer abstracts the physical storage medium, presenting a uniform block-based interface to upper-layer software. It manages data placement, scheduling, and I/O optimizations such as request merging and reordering. Linux’s block I/O subsystem includes components like BIO, I/O schedulers, and the multi-queue block IO queuing mechanism (blk-mq) for high-performance devices like NVMe SSDs. Note that the Block I/O layer can be bypassed via DAX (Direct Access) mode, allowing applications to directly access memory-mapped storage without traditional block I/O overhead. In addtion, the block layer can be integrated with storage virtualization techniques (e.g., Logical Volume Manager (LVM),  software RAID) to further enhances flexibility and resilience.

\smallskip
\noindent
\textbf{File Systems (FS).} File systems implement a set of file-related system calls (e.g., \texttt{open}, \texttt{read}, \texttt{write}, \texttt{close}) to provide the file and directory abstraction to  user applications. Internally, it formats the storage device into a set of data blocks and  metadata structures (e.g., bitmaps, inodes) to ensure efficient management and access of data on device. 
Traditional file systems (e.g., Ext4~\cite{nordvik2022ext4}, BtrFS~\cite{rodeh2013btrfs}) have been designed for managing data on HDDs. More recently, new file systems have been proposed for new devices. 
For example, F2FS~\cite{lee2015f2fs} is a special file system carefully designed for SSDs. Similarly, For persistent memory (PM), specialized file systems like NOVA~\cite{NOVA} and PMFS~\cite{PMFS} optimize performance by leveraging PM’s byte-addressability. 
Moreover, traditional file systems including EXT4 and XFS have been extended with DAX (Direct Access) support, leading to EXT4-DAX and XFS-DAX. The DAX mode eliminates the page cache and bypasses the block layer, allowing applications to directly access persistent memory at near-DRAM speeds. 
In general, different file systems mainly differ in terms of their on-drive data structures and access methods, while their interfaces are largely the same following the POSIX standard~\cite{posix}. 
Besides the basic file management, file systems may implement additional features such as journaling, copy-on-write (CoW),  deduplication, snapshots, encryption, and compression, etc. to enhance data integrity and security.

\subsection{User-Level  Management (UL): Enabling Various Applications  at Scale}

The user-level storage management software  enables various application scenarios (e.g., AI/ML applications, blockchain transactions). In particular, distributed storage systems (e.g., Lustre~\cite{Lustre}, HDFS~\cite{hdfs}, Ceph~\cite{redhat-ceph}) are designed to manage data at scale~\cite{Lustre,redhat-ceph,BeeGFS,google-colossus,orangefs,DAOS,Gluster,Swift}. They typically consist of a cluster of server nodes with different functionalities.
For example, Lustre is a distributed parallel file system   widely used in  high performance computing (HPC)  centers~\cite{Lustre,top500}. A Lustre cluster may  include a  MGS (Management Server) node to manage and store cluster-level configuration information, a  MDS (Metadata Server) node to manage and store the metadata,  and many {OSS} (Object Storage Server) nodes to manage and store the actual user data as objects and handle  I/O requests. Similarly,  Ceph is a distributed storage system designed to be highly scalable and fault-tolerant~\cite{redhat-ceph}, which typically consists of one monitor/manager node (MON/MGR) and many OSD (Object Storage Daemon) server nodes for user data.
 Clients may directly interact with multiple storage servers in the cluster for accessing their data, while metadata operations such as file indexing and directory management are typically handled by the metadata servers in the cluster.


Besides the basic functionalities which are largely similar, different DSS may introduce system-specific techniques to provide unique features.
For example, 
 Ceph storage system organizes objects   in a logical concept named \textit{pool}.
For better object management, objects in a Ceph pool are further devided by another logical group named placement group (PG).
PGs reside on one or more OSD devices and can be overlapped on OSDs.
On top of Ceph's object store service (RADOS), Ceph also integates object gateway (RGW), block device service (RBD) and file storage service (CephFS). 
In addition, to ensure fault tolerance, DSS typically include dedicated checking, recovery, and failure mitigation components.
For example, Lustre includes a  fault-tolerance component call \textit{LFSCK} to check and fix potential corruptions~\cite{LFSCK}. Similarly, Ceph supports  erasure coding (EC)  to provide redundancy of user data with low storage overhead. As of this writing, Ceph supports multiple EC plugins including  Reed-Solomon (RS) codes, Clay codes, etc.  via third-party libraries~\cite{RSCode,isa-ec-lib}.

Additionally, other storage-related applications (e.g., databases, blockchains) may exist at this level, offering structured data management and high-level storage abstractions for various application needs. Many of such storage applications can be backed  by an underlying DSS for basic storage service.
Note that while the majority of the code is typically implemented at the user-level, the storage applications may also include customized  OS kernel to improve performance.
For example, Lustre's \texttt{ldiskfs} backend is a variant of Linux Ext4 file system which modifies Ext4 and  relies on its extended attributes for metadata. Similarly, the latest version of Ceph has replaced the traditional \texttt{FileStore} backend with a customized \texttt{BlueStore} backend~\cite{ceph-bludstore} to reduce the latency at the OS kernel level.

\subsection{Summary}
In summary, a data storage system may include three main layers logically, including a storage device layer for storing data durability, an OS  layer for managing data on a single computer node (which may include multiple storage devices locally), and a user-level storage management layer for handling requests from the end users and various applications directly (Figure~\ref{fig:stack}).  
With the background of   storage system architecture described above,  we introduce a few representative works on analyzing and improving the fault tolerance of storage systems in the next section (\S\ref{sec:frameworks}).

\section{Fault Tolerance Analysis for Data Storage Systems}
\label{sec:frameworks}

     \begin{wrapfigure}{r}{0.45\textwidth}
      \centering
      \includegraphics[width=2.2in]{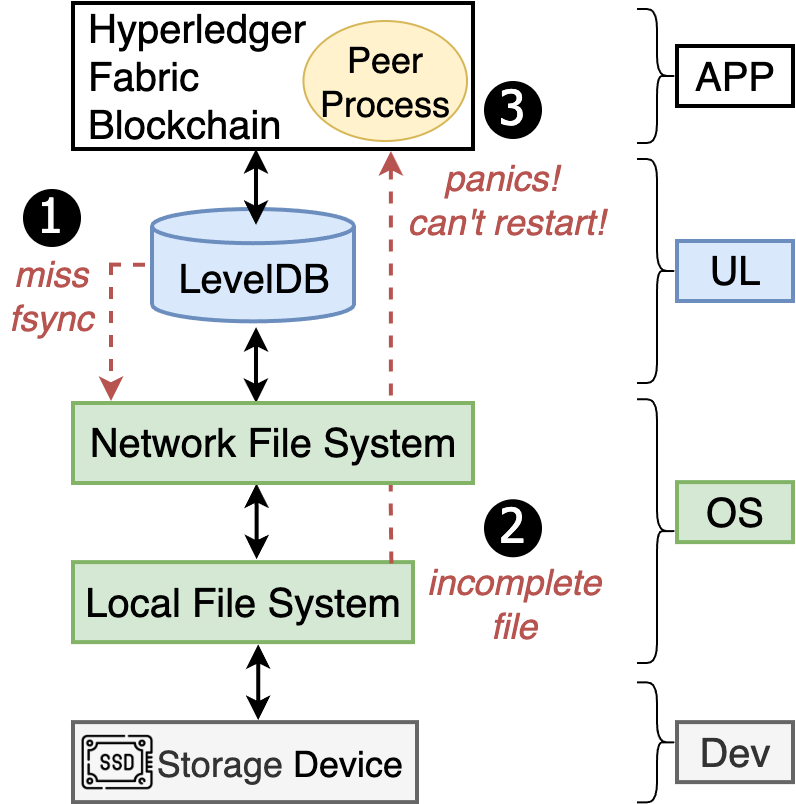}
  \caption{{\bf A Real-World Bug Case in Hyperledger Fabric Blockchain}. \textit{A peer process in the blockchain panicked and failed to restart}.}
     \vspace{-0.1in}
      \label{fig:bugcase}
    \end{wrapfigure}

 \smallskip
\noindent
\textbf{A Real-World Bug Case.} Given the complexity of data storage systems as introduced in \S\ref{sec:intro}, 
achieving end-to-end fault tolerance is challenging. Essentially,  applications and system layers do not live in isolation; there are  implicit dependencies across components in the  ecosystem (Figure~\ref{fig:stack}), which makes ensuring data integrity tricky under fault.
As one concrete example,  Figure~\ref{fig:bugcase} shows a real-world bug case~\cite{FAB-18304} occurred in Hyperledger Fabric Blockchain~\cite{androulaki2018hyperledger}. In this example,  the blockchain system accesses its state information stored in the  state database (i.e., LevelDB). The database  runs  on a Network File System (NFS) backed by a local file system and relies on the file systems to access data on the storage device, which forms a layered architecture (i.e., \texttt{Dev - OS - UL} as discussed in Figure~\ref{fig:stack} in \S\ref{sec:intro}).  
The root cause of the bug case lies in the LevelDB: it   misses an \texttt{fsync} system call when updating its internal metadata (\circled{1}), which may lead to an incomplete manifestation file (\circled{2}) in the file systems  in corner cases (e.g., when the capacity of the storage device is near full). When the bug is triggered, the peer process in the  blockchain system may panic and fail to restart (\circled{3}). The issue was labeled with the \textit{Highest} priority but it took five months to resolve~\cite{FAB-18304}, largely due to the complexity.
Similar issues caused by such cross-layer dependencies have led to widespread  failures of other blockchain systems in practice~\cite{aws2021feb,datta2022,prathap2021,pearson2021,blockchain2022}.  

     \begin{wrapfigure}{r}{0.55\textwidth}
      \centering
      \includegraphics[width=2.8in]{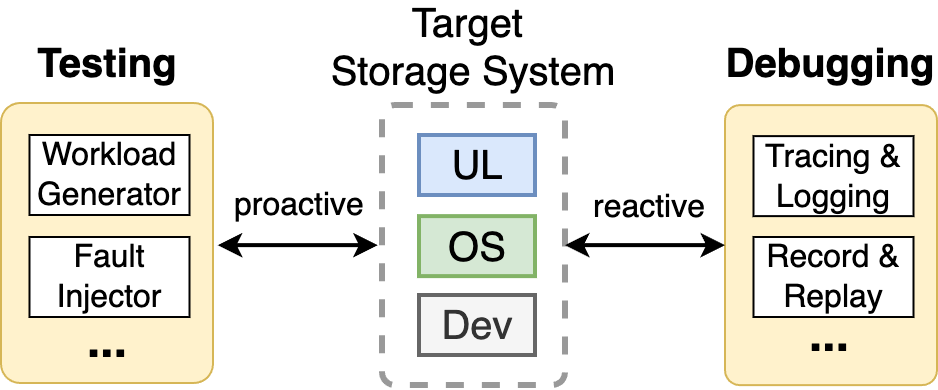}
  \caption{{\bf Testing \& Debugging Techniques are Essential for Enhancing the Fault Tolerance of  Target Systems}.}
     \vspace{-0.1in}
      \label{fig:testdebug}
    \end{wrapfigure}

 \smallskip
\noindent
\textbf{What We Need.} Addressing the grand challenges require  research innovations and collective efforts from the communities. In particular, \textit{Testing} and \textit{Debugging} are two  essential and complementary approaches  for ensuring the fault tolerance of data storage systems in general, as illustrated in 
 Figure~\ref{fig:testdebug}. More specifically,  \textit{Testing} is a \emph{proactive} approach used to identify potential defects in target storage systems before they impact a production environment. This method typically involves generating various workload with different patterns (i.e., a \emph{Workload Generator}), injecting or simulating faults (i.e.,  a \emph{Fault Injector}), and stressing the system to evaluate its robustness under different failure scenarios. Techniques such as stress testing, fault injection, and endurance testing help uncover weaknesses in the target storage system stack including  hardware, software, and network configurations. By running extensive test cases, system testing can detect performance bottlenecks, hardware limitations, and software bugs early in the development cycle, reducing the risk of unexpected failures. However, despite its effectiveness, system testing cannot anticipate every possible failure, especially those arising from complex real-world interactions between system layers.

Complementarily, \textit{Debugging} is a \emph{reactive} approach aimed at diagnosing and resolving failures that have already occurred in a deployed system. This method involves generating and analyzing system logs, performance metrics, and memory dumps to pinpoint the root cause of an issue. Debugging tools can reconstruct system states, perform comparative analysis with functional systems, and detect anomalies using advanced diagnostic techniques (e.g.,  \emph{Record \& Replay}). Common debugging challenges include identifying failures caused by rare concurrency issues, software-hardware interactions, or hidden firmware bugs. Accurate root cause identification is critical, as misdiagnosis can lead to ineffective fixes, prolonged system downtime, and unnecessary hardware replacements. As storage systems become increasingly complex, debugging requires sophisticated tools and expertise to ensure timely and effective failure resolution.

In the rest of this section,
we first introduce a few representative techniques for analyzing the fault tolerance of widely used data storage systems (\S\ref{sec:analyze-dev}, \S\ref{sec:analyze-local}, and \S\ref{sec:analyze-dist}). We classify the works based on the target system layers involved in the analysis (i.e., \texttt{Dev - OS - UL}   in Figure~\ref{fig:stack}). 
Next, as one step toward addressing the open challenge,   we discusses potential solutions for analyzing the fault tolerance of data storage systems in an end-to-end manner (\S\ref{sec:together}).

\subsection{Analyzing Storage Devices}
\label{sec:analyze-dev}

\begin{figure}
\begin{center}
\includegraphics[width=4.8in]{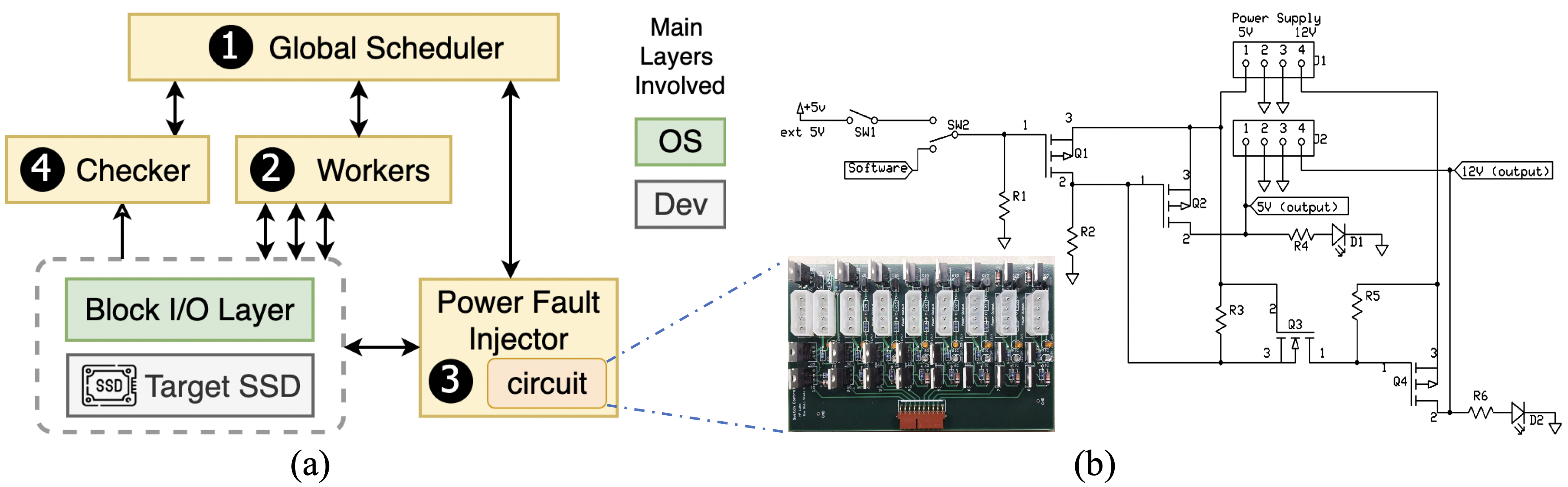}
\end{center}
\vspace{-0.2in}
\caption{{\bf A Fault Injection Framework for Analyzing SSDs under Fault}. \textit{(a) Workflow; (b) Fault injection circuit (adapted from ~\cite{our_fast13_ssd,our_tocs17}).}}
\label{fig:workflow}
\end{figure}

As discussed in \S\ref{sec:intro}, storage devices (e.g., HDD, SSD, PM) serve as the foundation of data storage systems. They are built with different technologies and  exhibit different characteristics, which may lead to different failure modes and require careful design of fault tolerance mechanisms at the upper layer. To understand the device-level behaviors, 
many researchers have conducted experiments and analysis on various devices at different granularity, including HDDs (e.g., ~\cite{bianca07,bairavasundaram2007analysis}), HDD-based RAID systems (e.g., ~\cite{gibsonthesis,krioukov2008parity}), raw flash memory
chips (e.g., ~\cite{ong_vlsi93,brand_irps93,suh_jssc95,belgal_irps02,kurata_vlsi06,Grupp09,powercut11,Grupp12,cai_date12,ycai_sigmetrics14}, flash-based SSDs (e.g.,~\cite{our_fast13_ssd,our_tocs17,bianca_fast16}), PMs (e.g.,~\cite{zhang2021study,yang2020empirical,gatla2023understanding}), etc. We briefly describe one representative work below, and refer the interested readers to~\cite{our_tocs17} for more details.

 Figure~\ref{fig:workflow} shows one  example of fault injection technique mainly focusing on the storage device   and a thin OS layer ~\cite{our_tocs17}. To understand the behaviors of SSDs under fault with minimal disturbance of the storage software, the target SSD is accessed as a raw block device through a thin software interface (i.e., the Block I/O layer). As shown in Figure~\ref{fig:workflow}a, the framework includes 
 four major
components: \textit{Global Scheduler},  \textit{Workers},  \textit{Power Fault Injector}, and  \textit{Checker}.  
The Global Scheduler (\circled{1}) coordinates the whole testing procedure including initializing the target SSD, selecting Workers (\circled{2}) to
apply carefully-designed workloads to the target for a predefined  period, sending a signal to the Power Fault Injector (\circled{3})  at a random time within the working period which turns off  the power
supply to the SSD accordingly, and invoking  the Checker (\circled{4})  to read the special records present on the restarted
SSD and check the correctness of the device state based on the record format. 
The Power Fault Injector component includes a dedicated circuit (Figure~\ref{fig:workflow}b) to enable efficient power fault injection with high fidelity.
The fault injection testing procedure is executed iteratively and all issues found are written to logs  for postmortem analysis. 
The framework has been applied to analyze dozens of SSDs from different vendors and exposed multiple failure modes of SSDs (e.g., shorn writes, serialization errors, bit corruption, metadata
corruption) that are different from traditional HDDs. 
The unique failure modes revealed in the experiments suggest the need of hardware-awareness in building fault-tolerance storage systems. 
For example, serialization errors exposed by the framework imply that traditional fault-tolerance solutions relying on the correct order of operations (e.g., write-ahead logging in databases or journaling file systems) might not be sufficient. Similarly, metadata corruptions and shorn writes imply that update-in-place to a sole copy of data is not enough to ensure fault tolerance. Interestingly, the number of errors observed might be affected by both the SSD device model and the OS kernel version of the  Block I/O layer, suggesting the dependency between the storage device and the OS kernel layers~\cite{our_tocs17}.
These findings have  raised the awareness and interest of power loss
protection and relevant fault tolerance issues (e.g., crash consistency) in general,
and the work has
inspired follow-up research on reliability across the storage stack in the community (e.g., ~\cite{our_atc19,alter2019ssd}).

\subsection{Analyzing Local Storage Software Stack}
\label{sec:analyze-local}

The local storage software stack serves as a critical intermediary between physical storage devices and user-level applications to manage data on a single computer (\S\ref{sec:intro}). 
Due to the prime importance, 
great efforts have been made to analyze and/or improve the reliability of  the local storage software, including Linux fault injection infrastructure~\cite{linuxFaultInjection}, regression test suites (e.g., xfstest~\cite{xfs_utilities}, e2fsprogs~\cite{e2fsprogs}),
configuration dependency analyzers (e.g., CONFD~\cite{fast23confd}), fuzzers (e.g., Syzkaller~\cite{syzkaller}), 
 record-replay tools for simulating faults and testing the crash consistency of  storage  software (e.g., ~\cite{our_osdi14_db,Vinter,leblanc2023chipmunk}), etc.
 We briefly describe two representative techniques below, and refer the interested readers to~\cite{linuxFaultInjection,our_fast18} for more details.

 \begin{figure*}[tb] 
\centering
\includegraphics[width=4.8in]{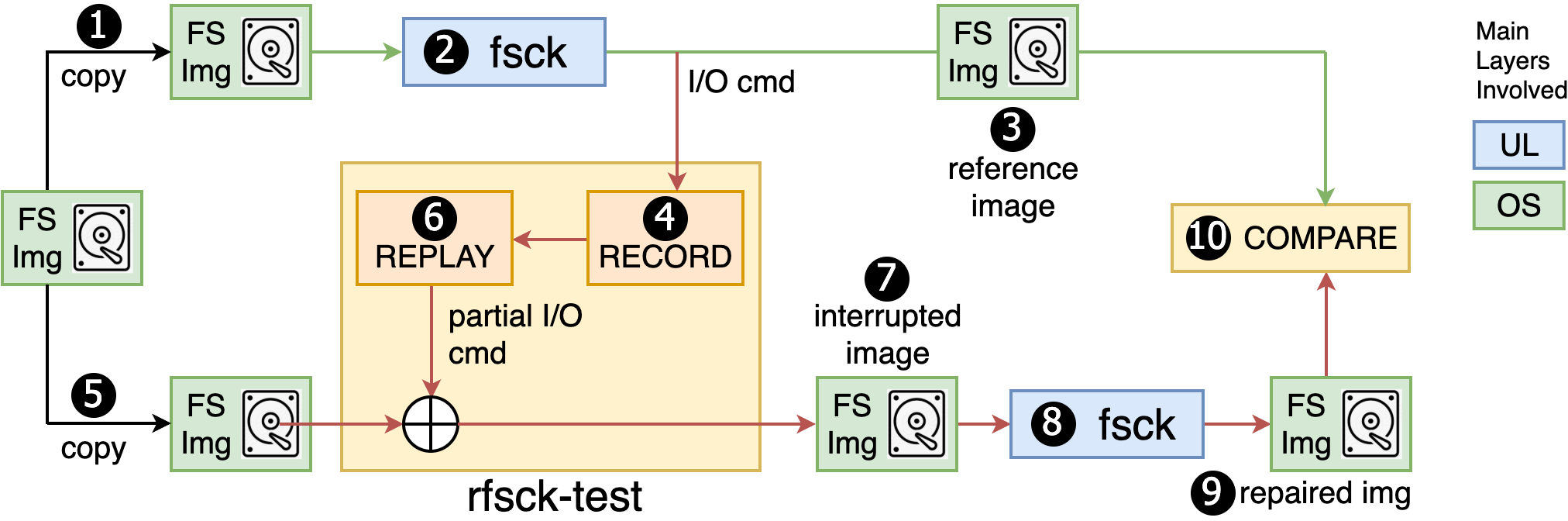}
\caption{
{\bf The RFSCK framework for analyzing the fault tolerance of file system checkers (adapted from \cite{our_fast18}).}
}
\label{fig:rfsckworkflow} 
\end{figure*}

\smallskip
\noindent
\textbf{Linux Fault Injection Infrastructure (LFI).} This is a fault injection framework introduced to  the Linux kernel  since version 2.6.20~\cite{linuxFaultInjection}. 
LFI can inject faults to the Linux kernel to simulate various issues (i.e., memory access errors) \cite{Subramanian-ICDE10-Corruption}. 
The implementation of this feature is located in the Linux source code under a dedicated path  (\texttt{lib/fault-inject.c}), 
and the LFI capabilities can be configured either at boot-time or during runtime\cite{linuxFaultInjection, Subramanian-ICDE10-Corruption}.
Roughly speaking, the LFI module reads the input parameters (e.g., interval, probability) from the \texttt{debugfs} string and stores the result in the fault attribute structure (\texttt{fault\_attr}). The core function (\texttt{should\_fail\_ex}) examines the capabilities, returns true if they are met, indicating a failure, or returns false if not.
Note that the LFI  feature is turned off by default, and activating it requires declaring specific directives in the kernel configuration and recompilation.

The LFI contains various  capabilities targeting different components in the Linux kernel~\cite{linuxFaultInjection}.
For example, \texttt{failslab} allows injecting  slab allocation failures in the kernel memory allocation functions (e.g., \texttt{kmalloc()} and \texttt{kmem\_cache\_alloc()})\cite{bonwick1994slab}. Similarly, \texttt{fail\_page\_alloc} allows
    injecting failures in memory page allocation, which can affect all the functions related to paging (e.g.,  \texttt{alloc\_pages()}, \texttt{get\_fee\_pages()}).

\begin{figure}[ht!]
    \centering
    \includegraphics[width=4.9in]{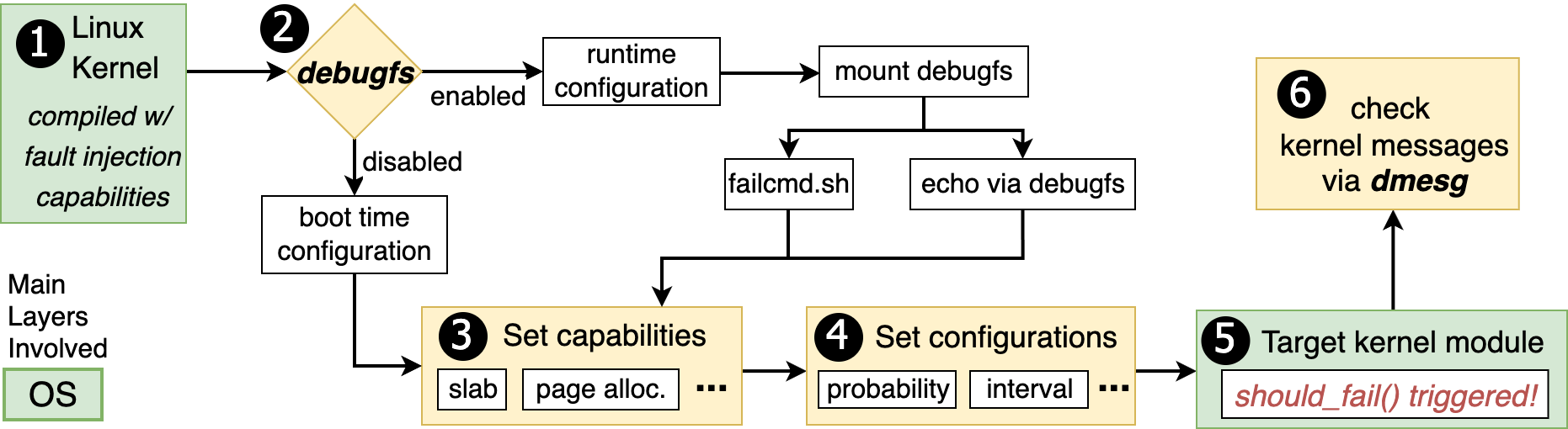}
    \caption{\bf An Example Workflow of Linux Fault Injection Framework.}
    \label{fig:LFI_flow}
\end{figure}

Figure \ref{fig:LFI_flow} illustrates an example  workflow of LFI to inject faults at a device driver.
 After configuring and compiling the kernel for fault injection, the next step is injecting faults into the target device through the \texttt{debugfs} interface \cite{debugfs}.
The fault injection is done by filling the fault capabilities directly through \texttt{debugfs}, where the target device driver that has been designed to be injected reads those capabilities and execute the injection action. LFI includes a predefined script (\texttt{failcmd.sh}) to inject the  fault, which is basically a bash script that utilizes \texttt{debugfs}.
When the injection value is received in the target device driver, the function (\texttt{should\_fail}) is triggered and causes the failure. Since LFI is at the OS level, errors can be directed to the user-level for further analysis. Finally, the \texttt{dmesg} tool collects kernel messages, including any errors  caused by the \texttt{should\_fail} function. 
Note that LFI is designed to be extendable to  include new capabilities. It can also  support additional fault models by adding the \texttt{should\_fail} function to different locations in the kernel source code. 

\smallskip
\noindent
\textbf{The RFSCK Framework.} This is one research prototype for injecting faults and analyzing the resilience of local file system checkers~\cite{our_fast18}. As shown in Figure~\ref{fig:rfsckworkflow},  there are ten main steps to test the fault tolerance of file system checkers systematically via RFSCK:
\circled{1}  the framework makes a copy of the test image which contains a corrupted file system;
\circled{2} the target checker (i.e.,~\texttt{fsck})  is executed 
to check and repair the original corruption on the copy of the test image;
\circled{3}  after ~\texttt{fsck} finishes normally in the previous step, 
 the resulting image is stored as the {\it reference image};
\circled{4} during the checking and repairing of~\texttt{fsck}, 
the fault injection tool 
 \texttt{rfsck-test} records the I/O commands generated by~\texttt{fsck} in a command history file (the basic mode);
\circled{5} the framework makes another copy of the original test image;
\circled{6}  \texttt{rfsck-test} replays partial commands recorded in step 4 to the new copy of the test image, 
which emulates the effect of an interrupted~\texttt{fsck};
\circled{7}  the image generated in step 6 is stored as the {\it  interrupted image};
\circled{8}  ~\texttt{fsck} is executed again on the interrupted image to fix any repairable issues;
\circled{9}  the image generated in step 8 is stored as the {\it  repaired image};
\circled{10}  finally, the framework compares the file system on the repaired image with that on the reference image 
to identify any mismatches.
Note that in step 8  ~\texttt{fsck} has been executed  without interruption, so a mismatch implies 
that there is some corruption which cannot be recovered by ~\texttt{fsck}. Also, besides the basic mode shown in the figure, RFSCK includes an advanced mode for testing file system checkers with logging support~\cite{our_fast18}. 
As of this writing,
the RFSCK framework has been applied to test  the checkers of multiple widely used file 
systems (i.e., ~\texttt{e2fsck}~\cite{e2fsprogs} for Ext-series file systems, ~\texttt{xfs-repair}~\cite{xfs_utilities} for XFS file system, \texttt{btrfs-fsck} for BtrFS file system, and \texttt{f2fs-fsck} for F2FS file system). The experimental results have demonstrated multiple vulnerabilities in the local file system layer  (e.g.,  the 
file system may be left in an uncorrectable state 
 if the 
repair is interrupted~\cite{our_tos18}).

\subsection{Analyzing Large-Scale  Storage Systems}
\label{sec:analyze-dist}

\begin{figure*}[t] 
    \centering
    \includegraphics[width=4.8in]{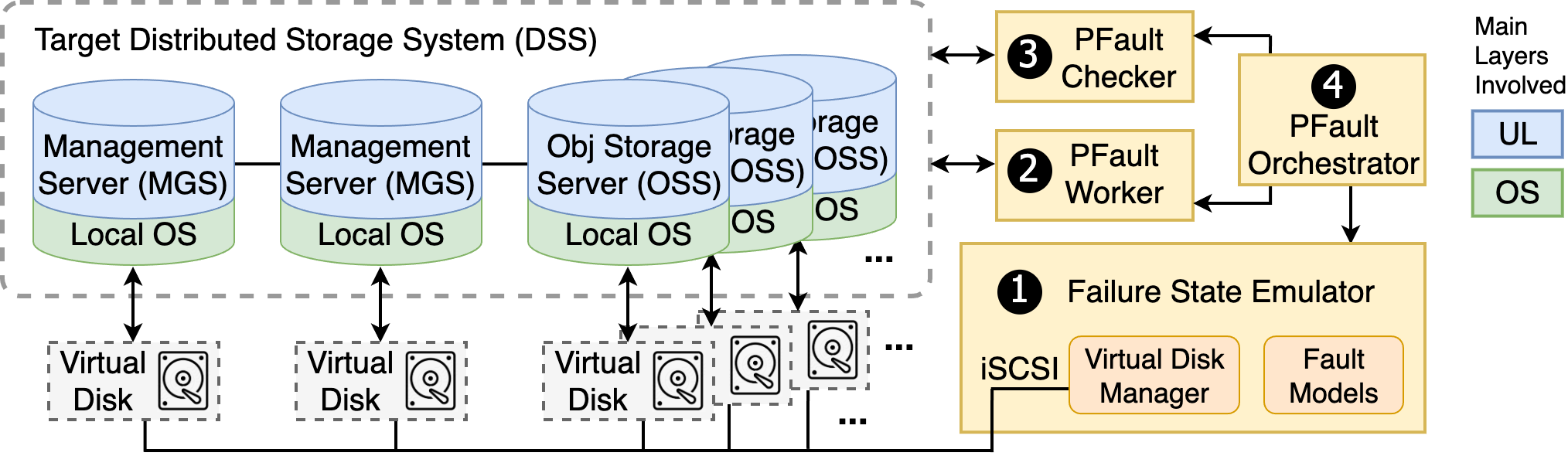}
    \vspace{-0.15in}
    \caption{
    {\bf The PFAULT framework for analyzing the fault tolerance of large-scale storage systems (adapted from \cite{our_ics18,han2022study}). }
     }
    \label{fig:pfault} 
    \end{figure*}

Enabled by local storage software stack,  large-scale storage systems manage resources across computer nodes  to support various services and application scenarios at scale (e.g., cloud object storage, distributed file systems,  blockchains)~\cite{Lustre,redhat-ceph,BeeGFS,gfs_sosp03,google-colossus,orangefs,DAOS,WekaIO,Gluster,Swift,hdfsmanual,cassandra,androulaki2018hyperledger,HadoopYarn,zookeeper}. 
Similar to other practical software systems, large-scale storage systems typically have built-in test suites to ensure their robustness (e.g., ~\cite{Lustre,BeeGFS}).  
In addition, many research prototypes have been proposed to understand the reliability of distributed systems  (e.g., ~\cite{ORCHESTRA,DOCTOR,NFTAPE,FIAT,ganesan2017redundancy,Jepsen,Yuan,Gunawi-NSDI11,Tanakorn-OSDI14-SAMC,scalecheck,Lu-SOSP19,alagappan2018protocol}). We briefly describe two representative frameworks below, and refer the interested readers to~\cite{our_ics18,awsfis} for more details.

\smallskip
\noindent
\textbf{The PFAULT Framework.} As one  example of analyzing the fault tolerance of large-scale storage systems,
we introduce a scalable fault injection framework called ~\pfault{}~\cite{our_ics18,han2022study}, which has been used to analyze the failure handling and recovery of multiple storage systems at scale~\cite{our_ics18,han2022study,han2020fingerprinting}.
As shown in Figure~\ref{fig:pfault},  ~\pfault{} includes  four major components: \textit{Failure State Emulator}, \textit{PFault Worker}, \textit{PFault Checker}, and \textit{PFault Orchestrator}. And it can be connected to the target distributed  storage system (which may include different types of  nodes as described in \S\ref{sec:intro}) via remote storage protocols (e.g., iSCSI~\cite{meth2003iscsi}).  
More specifically, 
the \textit{Failure State Emulator} (\circled{1}) is the key component responsible for injecting faults to trigger the fault tolerance  procedures of
 the target system.
It  mounts a set of virtual devices to the  storage nodes via iSCSI and forwards all device I/O commands to the backing files through the protocol, each of which represents the persistent state of a corresponding storage node.     
Moreover, the persistent states of storage nodes are manipulated by \textit{Failure State Emulator} to
emulate system failure states based on  workloads and a set of predefined fault models (e.g., \textit{Whole Device Failure}, \textit{Global Inconsistency}, \textit{Network Partitioning})
derived from real-world failure scenarios reported in the literature~\cite{Remzi-TOS08-Corruption,bairavasundaram2007analysis,nightingale_eurosys11,bianca07,Subramanian-ICDE10-Corruption,Ma-FAST15-RAIDShield,NP}.
Besides the \textit{Failure State Emulator}, the \textit{PFault Worker} (\circled{2})  launches workloads to exercise the target system and generate I/O 
operations; the \textit{PFault Checker} (\circled{3}) invokes the failure handling and recovery utilities (i.e., LFSCK for Lustre) of the target system as well as a set of verifiable workloads to examine the fault tolerance of the target system thoroughly; and the \textit{PFault Orchestrator}  (\circled{4}) component coordinates the overall workflow and collects the corresponding logs automatically for in-depth fault tolerance analysis. 
As of this writing, the \pfault{} framework has been applied to study two large-scale production storage systems including  Lustre~\cite{Lustre} and BeeGFS~\cite{BeeGFS}. The experimental results have exposed multiple cases where the target systems' fault tolerance guarantee is imperfect (e.g., the recovery procedure itself may hang, fail abruptly, or trigger kernel panics when scanning the storage nodes in the cluster~\cite{han2022study}).

\smallskip
\noindent
\textbf{AWS  Fault Injection Simulator (FIS) Service.}
Since the publication of PFAULT~\cite{our_ics18}, many efforts have been made to improve further the fault tolerance of distributed systems.
 Notably, Amazon recognizes the importance of fault injection  and  commercializes a service called Fault Injection Simulator (FIS)~\cite{awsfis}. The FIS shares similar design goals and principles as PFAULT, but extends the target to general distributed systems. To achieve the generality, it allows integration with third-party utility programs (e.g., stress-ng~\cite{stressng}) to conduct comprehensive testing and measurement.
 On the other hand, since FIS relies on   utility programs running
 \textit{inside} the target system to simulate faults, it might potentially change the target system and affect the  fidelity. 
Therefore, FIS is more suitable for testing user-level applications instead of testing full system stack.
Also, FIS is a commercial service relying on other AWS services (e.g., EC2, CloudWatch) to work, which might not be applicable to on-premise systems or non-commercial use cases with limited budget. Therefore, additional  efforts are probably needed to make the comprehensive fault tolerance analysis capability generally available.

\subsection{Putting It All Together: Towards Full-Stack Fault Tolerance Analysis}
\label{sec:together}

The frameworks introduced in previous sections ( \S\ref{sec:analyze-dev}, \S\ref{sec:analyze-local}, \S\ref{sec:analyze-dist}) are representative    techniques for analyzing the fault tolerance of different target storage systems.  While they are excellent for their original purposes, they are still insufficient to address the end-to-end fault tolerance challenge because  they only focus on a subset of all major layers in the data storage system hierarchy (Figure~\S\ref{fig:stack}), and thus cannot capture  the inherently dependencies across all major layers. 
To address the limitation, researchers have looked into full-stack approaches to improve the end-to-end coverage. We introduce a few efforts along this direction below.

Notably, VINTER~\cite{Vinter} is a framework  to support full-system testing of PM-based storage systems. By leveraging virtual machine (VM) technology, VINTER is designed to host a complete software stack. VINTER  has been applied to test PM-based file systems in the Linux kernel (e.g., PMFS~\cite{PMFS}) and has helped exposed multiple crash-consistency bugs that affect the fault tolerance of target systems~\cite{Vinter}.  
Unfortunately, 
while the VINTER approach is promising, our experiments found that it is  not scalable enough. There are multiple limitations based on our analysis: 
First of all, VINTER only supports minimal-built OS, which makes it incompatible to most application scenarios which depends on important system libraries (e.g., PMDK for PM application development).
Second, VINTER incurs significant overhead even under  a small set of writes from the applications (e.g., for a  workload with 256*20 bytes of writes, VINTER cannot finish the testing within 12 hours), which suggests that VINTER cannot be applied to real-world scenarios where write operations are common and write sizes are typically larger than a few KBs.
Third, VINTER has little  support for debugging the fault tolerance issues triggered and pinpointing the root causes.

\begin{figure*}[htbp]
	\centerline{\includegraphics[width=5in]{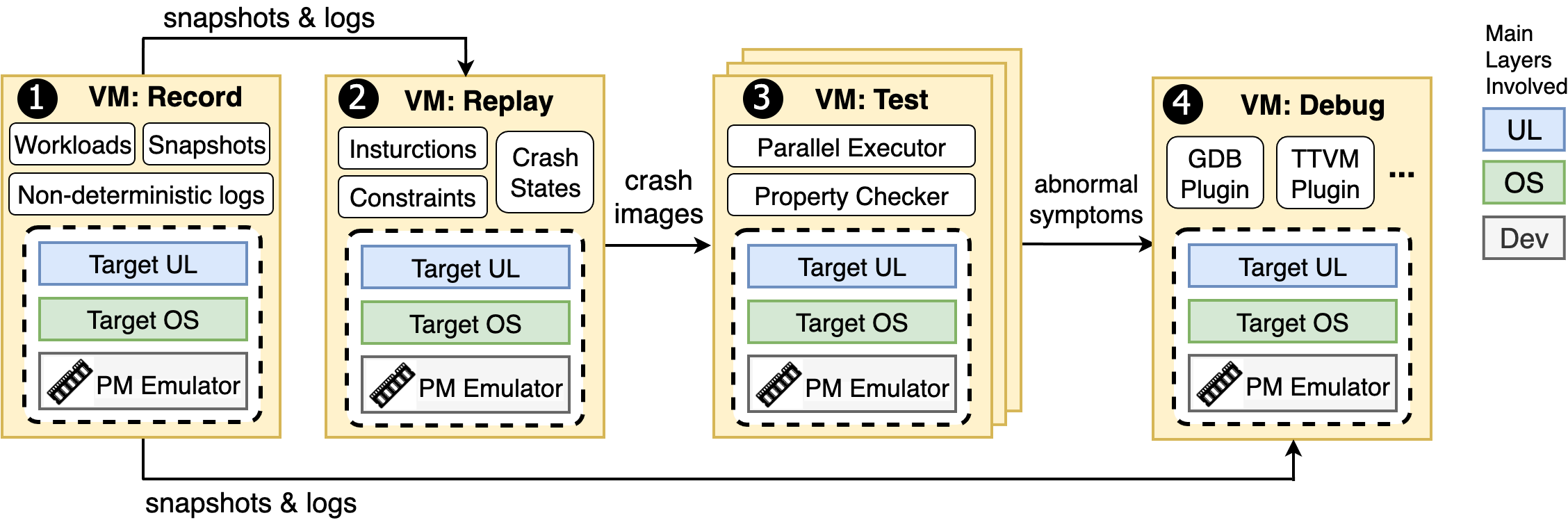}}
	\vspace{-0.1in}
	\caption{{\bf A VM-based Framework  (VFAULT) for Analyzing the Full Storage System Stack on PM.}}
	\label{fig:design}
	\vspace{-0.1in}

\end{figure*}

Inspired by VINTER as well as the key observations on its limitations, we 
propose a scalable VM-based framework called VFAULT for analyzing the fault tolerance of the entire storage system stack on PM. 
As shown in  Figure~\ref{fig:design}, the framework leverages customized VMs to emulate storage devices (e.g., PM) and host the entire software stack from OS kernel to applications. It supports a set of critical features that are important for scalable fault tolerance analysis (e.g., record and replay, tracing and debugging, parallel crash state generation and testing). We elaborate on the main steps  below.

The VFAULT workflow begins with an initial input system image that encompasses the full system software stack (e.g., the OS kernel,  applications and required libraries, and all environmental configurations).
The guest OS image is initiated with root privileges in the VM. 
After booting the system, the framework enters the recording phase (i.e., \circled{1} ``VM: Record''). It captures a snapshot as the original snapshot, comprising emulated device status (e.g., PM) and CPU register information. 
Moreover, the framework runs testing workloads on the pre-generated snapshot and leverages an architecture-neutral dynamic analysis tool called PANDA~\cite{PANDA} to record the execution of the whole system stack under the testing workloads,  which  captures relevant system calls and critical device commands/instructions (e.g., PM instructions). The recording phase generates two types of files (i.e., \textit{snapshots} and \textit{non-determinism logs}) to be utilized for testing and debugging support in the following steps.

Next, in the replay phase (i.e., \circled{2} ``VM: Replay''), the framework extracts storage commands and/or memory instructions (e.g., PM instructions) from the recorded executions. The framework  generates  a series of crash images based on the extracted instructions/commands and corresponding constraints (e.g., specifications of PM programming models, customized rules). These crash images represent the  persistent states of the target systems under various fault scenarios which can be used for parallel  testing and  fault tolerance analysis. 

In the testing phase (i.e., \circled{3} ``VM:  Test), the framework checks the recovery of the target system on crash states, and validates if there are any crash consistency issues (i.e., system fails to recover to a consistent state from the crash images). Given the complexity of the storage system and the variety of possible crash states,  the framework runs concurrent VMs to generate crash images in parallel
 and test the recoverability and crash consistency in parallel too. In this way, framework can  improve the scalability with reasonable tradeoffs (e.g., VM resources versus scalability). 
In addition, the framework is integrated with both classic debugging tools (e.g., GDB~\cite{gdb},    time-traveling virtual machine (TTVM)~\cite{king2005debugging}) to support pinpointing the root causes of crash consistency issues exposed (i.e., \circled{4} ``VM: Debug'').

Note that the VFAULT framework is designed to be extensible to support third-party debugging tools, customized crash consistency rules, and future programming models.
One key technique enabling these features is  the snapshot mechanism. 
The framework generates   a privileged snapshot immediately after the guest OS boots. This snapshot has two critical components:
(1) Memory Status, which includes all data residing in the current emulated device;
(2)  CPU States, which  includes the current register values for the emulated CPU.
Subsequently, the snapshot is incorporated into various phases of the framework's execution. For example, in the recording and tracing phase, we load the pre-generated snapshot to support executing the workload to genrate critical memory operations (e.g., fence and cache line flush instructions on PM). Similarly, in the testing phase, all subsequent crash states are generated based on the updates to the original snapshot (and thus creating additional  snapshots for representing crash images).

At the time of this writing, a prototype of the VFAULT framework has been applied to test the crash consistency of full storage software stack under multiple application scenarios. Table~\ref{tab:fullsys} shows one set of  experimental results for testing PM-based systems. 
In this set of experiments, VFAULT emulated a 128MB PM as the storage device and configured it accordingly through the guest OS kernel in VM.
We evaluated six PM applications on top of the PM software stack, including \textit{B-tree, C-tree, RB-tree, Hashmap\_atomic, Hashmap\_tx} and \textit{Hashmap\_rp} (as listed in the first column of Table~\ref{tab:fullsys}).
All PM applications relied on the PMDK library and Ext4-DAX file system support. Following the generation of crash states, we employed the Ext4 file system checker \texttt{e2fsck} to double check any corrupted states at the file system level. Moreover, we manually examined the recovered data structures of the workloads to validate any potential corruptions. 

\begin{table*}[htbp]
\begin{tabular}{c|c|c|c|c|c}
\begin{tabular}[c]{@{}c@{}} {\bf Applications on}\\ {\bf Full Storage Stack} \end{tabular} & \begin{tabular}[c]{@{}c@{}} {\bf Cksum} \\  {\bf Err}\end{tabular} & \begin{tabular}[c]{@{}c@{}} {\bf J-Txn} \\  {\bf Err}\end{tabular} & \begin{tabular}[c]
 {@{}c@{}} {\bf Metadata} \\ {\bf Err} \end{tabular} & \begin{tabular}[c]{@{}c@{}} {\bf Umount}  \\{\bf Err} \end{tabular} & \begin{tabular}[c]{@{}c@{}} {\bf Watchdog} \\ {\bf Bug} \end{tabular} \\ \hline \hline
B-tree & 4 & 1   & 3 & 1 & 1 \\ \hline
C-tree & 29 & 4   & 1 & 1 & 1 \\ \hline
RB-tree & 17 & 2   & 1 & 0 & 1 \\ \hline
Hashmap\_atomic & 8 & 4  & 1 & 3 & 1 \\ \hline
Hashmap\_tx & 8 & 4   & 0 & 0 & 1 \\ \hline
Hashmap\_rp & 6 & 4  & 0 & 0 & 0 \\ \hline
\textbf{TOTAL}  & \textbf{72} & \textbf{19} & \textbf{6} & \textbf{5} & \textbf{5} \\
\end{tabular}
\vspace{-0.2cm}
	\caption{ {\bf Full-System Testing Results under Six Application Scenarios.} }
\label{tab:fullsys}
	\vspace{-0.1in}
\end{table*}

The experiments exposed multiple fault tolerance issues of the target PM system stack. As summarized  in Table~\ref{tab:fullsys}, we observed five different failure symptoms including checksum errors (``Cksum Err''), journal transaction corruptions (``J-Txn Err''), metadata corruptions (``Metadata Err''), unmount errors (``Umount Err''), and watchdog bugs reported within the guest OS (``Watchdog Bug''). Further analysis indicates that some fault tolerance issues may be caused by the interplay and dependency between PM library (PMDK) and file system components ( \texttt{e2fsck}), which further suggests the importance of the holistic approach for ensuring end-to-end fault tolerance in practice. Note that the current prototype of VFAULT focus on single-node storage system stack; additional research and engineering efforts are needed to extend it to large-scale distributed storage systems, which we leave as future work.

\section{Conclusion \& Future Work}
\label{sec:conclusion}

In this chapter, we have described the general architecture of data storage systems, which mainly includes three layers: storage devices (Dev), local storage software stack in the operating systems (OS), and user-level applications which may be  distributed at scale (UL). We have also discussed the design and implementation of multiple representative fault injection testing frameworks for individual storage systems. While these frameworks are excellent for their original design goals, they are still relatively limited from an end-to-end perspective because there are inherent dependencies across layers which may affect the end-to-end fault tolerance guarantees of data storage systems in practice. 
Given the complexity of real-world data storage systems, we believe this is  an open challenge which probably requires collective efforts from the communities.
As one step towards addressing the grand challenge, we presented a VM-based full-stack testing framework called VFAULT, which currently focuses on the single-node storage system stack. Additional research efforts are likely needed to extend the idea to analyze the fault tolerance of large-scale storage systems in an end-to-end manner, which we leave as future work.
We hope that the comprehensive description of data storage systems and representative solutions presented in this chapter  
can inspire follow-up research on
analyzing cross-layer dependencies and ensuring end-to-end fault tolerance for mission-critical data storage systems (e.g., distributed databases, blockchain storage) in general.
\section*{Acknowledgments}
\label{sec:ack}
\vspace{-0.1in}
The authors would like to thank the members of Data Storage Lab (DSL) for research discussions  and  experiments on multiple projects covered by this chapter, including Jinrui Cao, Om Rameshwar Gatla, Runzhou Han, Tabassum Mahmud, 
Chao Shi, Wei Xu, Varun S. Girimaji,
Joshua John,  Roop Kiran, and
Vidhya Mannathu Parambil. This material was partially supported by US National Science Foundation (NSF) under grants CNS-1566554/1855565 (CRII), CCF-1910747, CNS-1943204 (CAREER), CNS-2402858, and a Global Research Outreach (GRO) Award from Samsung Advanced Institute of Technology (SAIT) and Samsung Research America (SRA). Any opinions, findings, and conclusions or recommendations expressed in this material are those of the author(s) and do not necessarily reflect the views of the sponsors.



\begin{backmatter}

\begin{authordetails}
	
	
	\author{Mai Zheng $^{1}$, Duo Zhang$^{2}$, Ahmed Dajani$^{3}$}
	\address[1]{Iowa State University, Ames, United States}
	\address[2]{Iowa State University, Ames, United States}
        \address[3]{Iowa State University, Ames, United States}
	%
	%
	
	\IntechOpentext{\textcopyright\ \the\year{} The Author(s). License IntechOpen. This chapter is distributed under the terms of the Creative Commons Attribution License (http://creativecommons. org/licenses/by/3.0), which permits unrestricted use, distribution, and reproduction in any medium, provided the original work is properly cited.}
	
	
\end{authordetails}


\bibliographystyle{abbrv}
\bibliography{mz_reference.bib}

\end{backmatter}

\end{document}